\documentclass[prb,twocolumn,showpacs,amsmath,amssymb]{revtex4}
\usepackage{graphicx}
\usepackage{dcolumn}
\usepackage{bm}
\usepackage{gensymb}
\usepackage{multirow}

\begin{document}

\title{Quasi-two-dimensional non-collinear magnetism in the Mott insulator Sr$_2$F$_2$Fe$_2$OS$_2$}

\author{Liang L. Zhao$^{1}$, Shan Wu$^{2}$, Jiakui K. Wang$^{1}$, J. P. Hodges$^{3}$, C. Broholm$^{2,4}$, and E. Morosan$^{1}$}
\affiliation{$^1$Department of Physics and Astronomy, Rice University, Houston, TX 77005, USA\\
             $^2$Institute for Quantum Matter and Department of Physics and Astronomy, The Johns Hopkins University, Baltimore, MD 21218, USA\\
             $^3$Neutron Sciences Directorate, Instrument and Source Design Division, Oak Ridge National Laboratory, Oak Ridge, TN 37831, USA\\
             $^4$Neutron Sciences Directorate, Quantum Condensed Matter Division, Oak Ridge National Laboratory, Oak Ridge, TN 37831, USA}

\date{\today}

\begin{abstract}
The magnetism of Sr$_2$F$_2$Fe$_2$OS$_2$ was examined through neutron powder diffraction and thermodynamic and transport measurements. Quasi-two-dimensional magnetic order develops below $T_N$ = 106(2) K with an in-plane correlation length exceeding 310 \AA\ and an out-of-plane correlation length of just 17(3) \AA. The data are consistent with a two-{\bf k} structure with $\textbf{k}_1$ = (1/2, 0, 1/2) and $\textbf{k}_2$ = (0, 1/2, 1/2) and an ordered moment of 3.3(1) $\mu_B$ oriented along the in-plane components of {\bf k}. This structure is composed of orthogonal AFM chains intersecting at super-exchange mediating O sites. Density Functional Theory (DFT) also points to this structure and a narrower Fe 3$d$ band than for the iron pnictides from which electronic correlations produce a Mott insulator.
\end{abstract}


\maketitle

Quasi-two-dimensional magnetic materials near the metal-insulator transition have been the focus of intense research since the discovery of high $T_c$ superconductivity in layered  copper oxides. The discovery of superconductivity in the chemically distinct iron pnictides has brought attention to the potential for novel electronic phases in layered iron-based compounds at the local-to-itinerant moment boundary.  Here we show that Sr$_2$F$_2$Fe$_2$OS$_2$ is a Mott insulator with quasi-two-dimensional non-collinear magnetic order. The unusual magnetic structure features \textit{two} wave-vectors $\textbf{k}_1$ = (1/2, 0, 1/2) and $\textbf{k}_2$ = (0, 1/2, 1/2)  associated with perpendicular AFM spin chains that share super-exchange mediating oxygen sites.

Sr$_2$F$_2$Fe$_2$OS$_2$  is structurally related to La$_2$O$_3$Fe$_2$X$_2$ (X = S, Se, ``2322") which was first reported by Mayer \emph{et al.}\cite{LaFeX2322_mayer} This is a a layered, tetragonal structure consisting of stacked [La$_2$O$_2$]$^{2+}$ and [Fe$_2$OX$_2$]$^{2-}$ sheets, composed of edge-sharing La$_4$O tetrahedra and face-sharing FeO$_2$X$_4$ octahedra, respectively. The Fe$_2$OX$_2$ sublattice forms a checkerboard spin lattice, with O and X mediated Fe-Fe super-exchange interactions. Insulating behavior and long range antiferromagnetic (AFM) ordering at 105 K (X = S) or 93 K (X = Se) was reported in these systems. Density Functional Theory (DFT) indicates these compounds are Mott insulators.\cite{LaFeX2322_PRL}

Very recently, the isostructural compounds La$_2$O$_3$Co$_2$Se$_2$,\cite{LaCoSe2322_JACS, LaCoSe2322_Fuwa} R$_2$O$_3$Mn$_2$Se$_2$ (R = La, Ce, Pr),\cite{LaMnSe2322_ni, LaMnSe2322_BaMnSe22122_XHChen, RTSe2322} R$_2$O$_3$Fe$_2$S$_2$ (R = Ce, Pr) \cite{RFeS2322} and R$_2$O$_3$Fe$_2$Se$_2$ (R = La-Sm) \cite{RTSe2322,RFeSe2322} were reported. All  ``2322" materials known so far are insulators or semiconductors with AFM order.\cite{LaCoSe2322_JACS} While no band structure calculations were reported for the Mn compounds,\cite{LaMnSe2322_ni, LaMnSe2322_BaMnSe22122_XHChen, RTSe2322} transport and magnetic properties similar to those of La$_2$O$_3$Fe$_2$S$_2$ indicate a Mott-localization picture may be applicable.

By replacing [La$_2$O$_2$]$^{2+}$ sheets with different 2+ valence spacer layers, several other related compounds were synthesized, including A$_2$T$_2$OX$_2$ with A = SrF, BaF (``22212")\cite{AFeX22122} or Na (``2212"),\cite{Na2Fe2OSe2} and T = Mn, Fe and X = S, Se.\cite{AFeX22122}  For a view into electronic correlations in this family of materials, we examine the magnetic structure of Sr$_2$F$_2$Fe$_2$OS$_2$ finding it distinct from anything seen in the copper oxides or the iron pnictides.

\begin{figure}
  \includegraphics[width=\columnwidth]{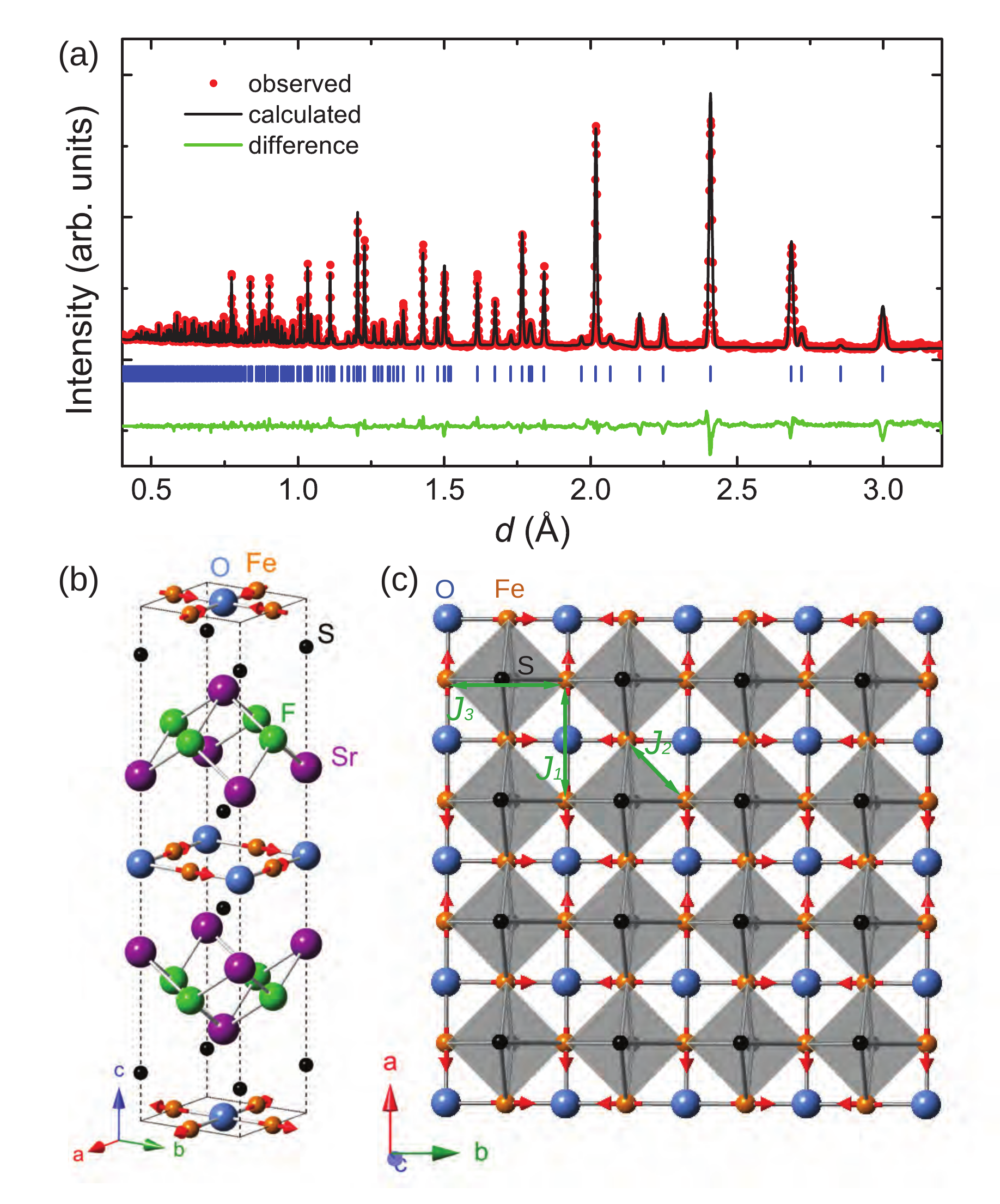}\\
  \caption{Sr$_2$F$_2$Fe$_2$OS$_2$ (a) neutron diffraction pattern for $T$ = 300 K (red - measured, black - calculated, green - difference between the measured and calculated patterns, blue vertical lines - calculated nuclear peak positions) and (b) the unit cell, with the red arrows indicating the Fe spins orientation in the AFM state, and (c) the $ab$ plane view of the Fe$_2$OS$_2$ layer.}\label{Neutron1}
\end{figure}

\begin{figure}
  \includegraphics[width=\columnwidth]{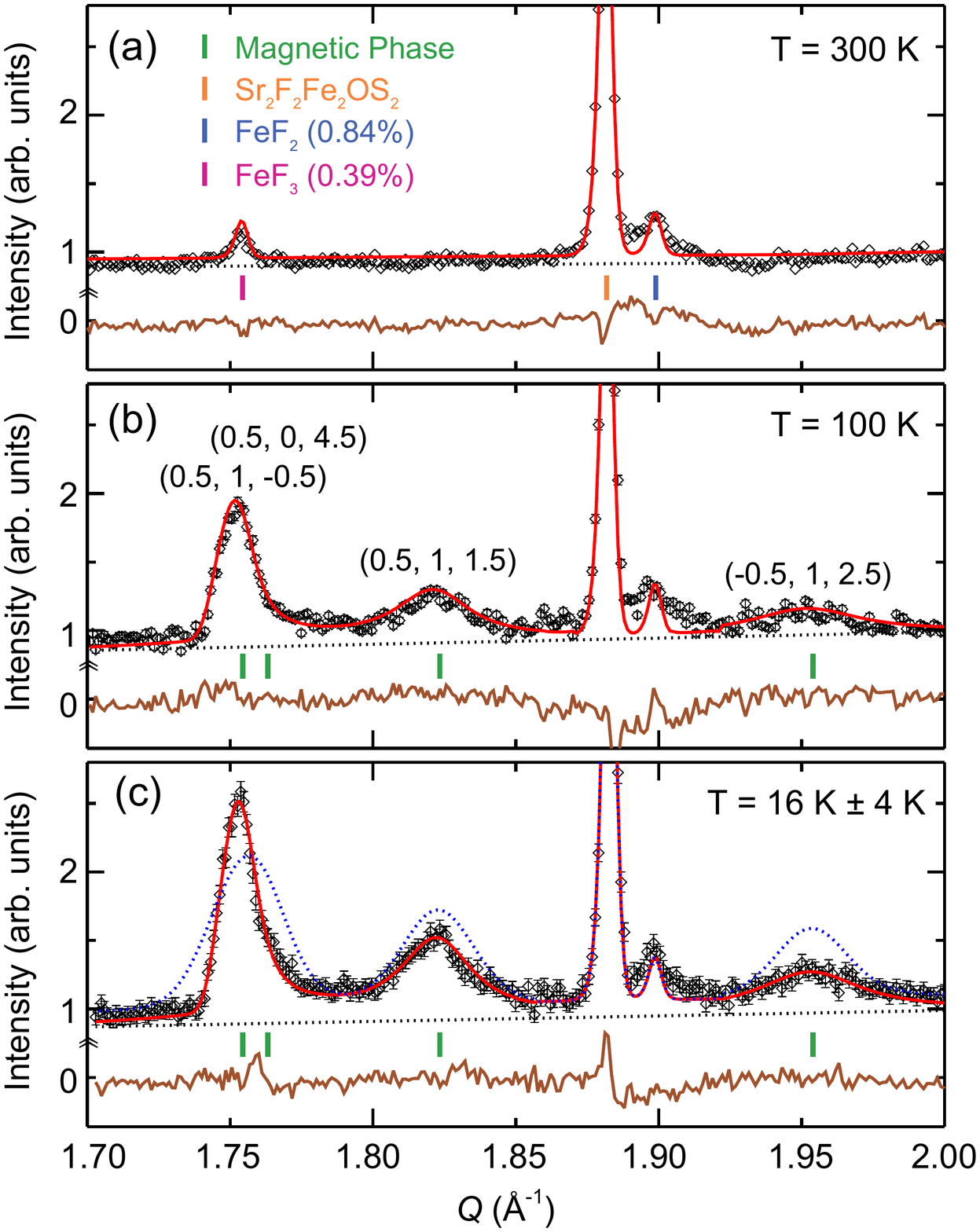}\\
  \caption{Neutron diffraction patterns for $T$ = (a) 300 K, (b) 100 K and (c) 12 K and 20 K (averaged), showing the development of magnetic peaks below $T_N$ = 106(2) K  (green tick marks). The magnetic peaks are refined with a Warren-like line shape (red lines, see text) that consistently accounts for the magnetic diffraction data. The fitting background and the difference between the measured and calculated patterns are shown in black dotted lines and brown solid lines, respectively. The blue dotted line represents the best fit profile assuming an isotropic magnetic correlation length, which is however inconsistent with the experimental data.}\label{Neutron2}
\end{figure}

Polycrystalline Sr$_2$F$_2$Fe$_2$OS$_2$ was synthesized by solid state reaction, using SrF$_2$, SrO, Fe and S powders as starting materials. The mixture was pelletized and annealed in vacuum at temperatures between 800 $\celsius$ and 850 $\celsius$.  Neutron powder diffraction data were collected for temperatures between 12 K and 300 K on the time-of-flight POWGEN powder diffractometer at the Spallation Neutron Source (SNS) at Oak Ridge National Laboratory. In order to focus on the chemical and magnetic structures respectively, the data were acquired using two different instrumental configurations with wavelength bands of 1.066 $\AA$--2.132 $\AA$ (Fig. \ref{Neutron1}a) and 4.264 $\AA$ --5.33 $\AA$ (Fig. \ref{Neutron2}). The room temperature diffraction pattern (red points, Fig. \ref{Neutron1}a) was refined to a tetragonal structure (Fig. \ref{Neutron1}b) using \textsc{fullprof}.\cite{FullProf} The lattice parameters $a$ = 4.0362(5) $\AA$ and $c$ = 17.9915(1) $\AA$ are consistent with the previously reported values.\cite{AFeX22122} The refined atomic positions and isotropic thermal factors are provided as supplemental material. \cite{SuppMat} Minute amounts of impurity phases (0.84\% FeF$_2$ and 0.39\% FeF$_3$) were also detected.

For temperature below $T_N$ = 106(2) K, additional peaks (green tick marks, Fig. \ref{Neutron2}b and c) appear in the neutron diffraction pattern. These correspond to a propagation vector \textbf{k} = (1/2, 0, 1/2) and will be seen to arise from magnetic scattering. Two unusual aspects of the magnetic diffraction pattern will be shown to define the magnetic structure: (1) The observable magnetic peak with the shortest wave vector is at $Q=1.75$~\AA$^{-1}$ and is associated with indexes (0.5, 0, 4.5) and (0.5, 1, --0.5). This points to a longitudinally modulated magnetic structure where the polarization factor extinguishes peaks at wave vectors that do not have an adequate component transverse to the in-plane direction of unit cell doubling (here the \textbf{a}-direction); (2) The (0.5, 1, --0.5) magnetic Bragg peak stands out by a sharp leading edge which is not apparent at (0.5, 1, 1.5). This indicates strong correlation length anisotropy with much longer correlations in the \textbf{a-b} plane than along \textbf{c}.

For a systematic analysis of possible magnetic structures we used representation analysis and Rietveld refinement as implemented in \textsc{sarah}\cite{SARAh} and \textsc{fullprof}.\cite{FullProf} For the $I/4mmm$ symmetry with a single propagation vector of \textbf{k} = (1/2, 0, 1/2), the iron sites are divided into two independent sets: those forming Fe-O-Fe chains along the in-plane \textbf{a}-component of \textbf{k} = (1/2, 0, 1/2) and those forming Fe-S$_2$-Fe chains along \textbf{a} (Fig.~\ref{Neutron1}c). Using Kovalev notation,\cite{Kovalev} the representation $\Gamma_{mag}$ associated with such magnetic structures decomposes into two irreducible representations (irreps), $\Gamma_{1}$ and $\Gamma_{3}$ in the form of $\Gamma_{mag}=\Gamma_{1}+2\Gamma_{3}$, with all three basis vectors projected. Given the observations above which are quantitatively born out by the Rietveld analysis, $\Gamma_{3}$ with moments along the in-plane component of \textbf{k} is the only irrep that is consistent with the data.

As the refinement based on the model of isotropic correlation length (blue dotted line, Fig. \ref{Neutron2}c) failed to account for the experimental data ($\chi^{2}$ = 7.496), the correlation length anisotropy was quantified by replacing in the Rietveld analysis of magnetic diffraction the spherical average of $\delta^3 ({\bf Q-\tau})$ by the spherical average of the following normalized anisotropic correlation function (Warren-like function\cite{Warren}):
\begin{equation*}
    {\cal S}(\textbf{Q})=\frac{1}{2\pi\sigma_{\perp}^2}\exp(-\frac{1}{2}(\frac{|{\bf Q-\tau}|_{\perp}}{\sigma_{\perp}})^2)\frac{\xi_{c}}{\pi}\frac{1}{1+(\xi_c|{\bf Q-\tau} |_{\parallel})^2}
\end{equation*}
The gaussian and lorentzian respectively describe long range in-plane correlations and short range correlations along \textbf{c}.
The polarization factor associated with the $\Gamma_{3}$ structure was implemented within the spherical average. The corresponding fit shown in Fig.~\ref{Neutron2} provides an excellent account of the magnetic diffraction data ($\chi^{2}$ = 1.033). The in-plane peak width $\sigma_{\perp}=0.0036~\AA^{-1}$ is near the resolution width extracted from the nearby nuclear peak: $\sigma_{res}=0.0023$~\AA$^{-1}$, indicating the in-plane correlation length exceeds 310~\AA. By contrast, the correlation length along \textbf{c}, $\xi_{\perp} = 17(3) \AA$, is indistinguishable from the lattice parameter c, so the low temperature magnetic order can rightfully be characterized as quasi-two-dimensional. These results are consistent with the initial phenomenological observations.

\begin{figure}
  \includegraphics[width=\columnwidth]{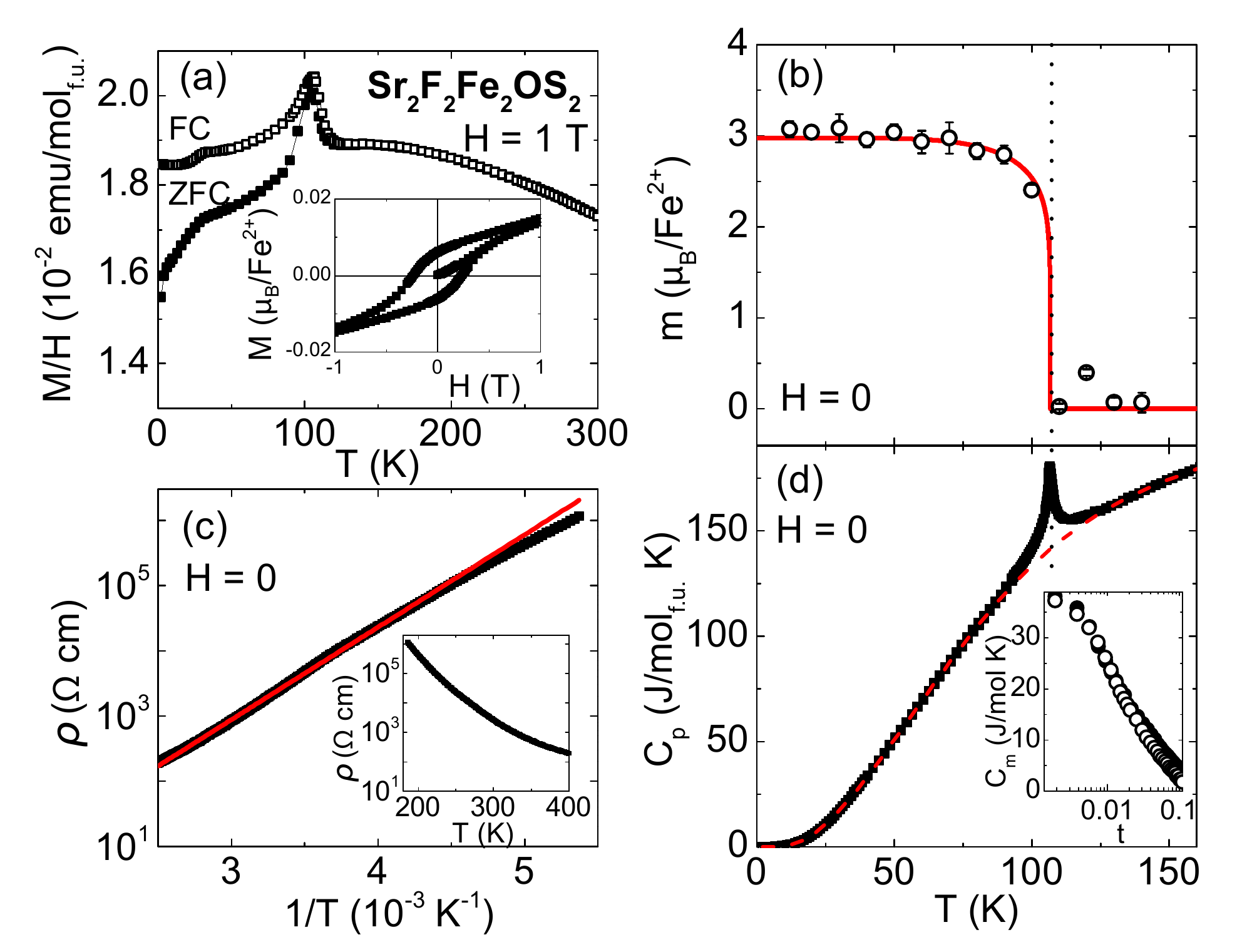}\\
  \caption{(a) DC magnetization measured for $H$ = 1 T. The ZFC and FC data are shown as solid and open symbols, respectively; inset: the hysteresis loop $M(H)$ for $T$ = 2 K. (b) The $T$-dependent staggered magnetization obtained by fitting diffraction data. The solid line shows the Onsager solution of the 2D square lattice Ising model. (c) semi-log plot of resistivity vs. $1/T$  (symbols), with a linear fit that gives $E_g$ = 0.28 eV. The raw $\rho(T)$ data are shown in the inset; (d) zero field heat capacity. After subtracting the phonon contribution (dashed line), the magnetic heat capacity $C_{m}(T)$ is plotted vs. $t=|T-T_{N}|/T_{N}$ (inset), showing logarithmic divergence near $T_N$ for both $T<T_{N}$ (solid) and $T>T_{N}$ (open).}\label{MCR}
\end{figure}

A well-known ambiguity presents itself in interpreting the results. The Fe-O-Fe and Fe-S$_2$-Fe chains associated with $\textbf{k}_1=(1/2, 0, 1/2)$ have identical, real magnetic structure factors for allowed magnetic Bragg peaks in the collinear $\Gamma_3$ spin structure. The magnetic diffraction intensity from a single wave vector domain thus depends only on the sum of the ordered moment on the two sites. By calibrating the magnetic diffraction intensity against the nuclear diffraction intensity in the conventional fashion, the magnitude of the total moment comes out to be 4.7(1) $\mu_B$. It is not possible, on the basis of diffraction alone, to distinguish the collinear structure from a non-collinear two-\textbf{k} structure where perpendicular Fe-O-Fe chains are modulated antiferromagnetically (Fig.~\ref{Neutron1}c). However, while the collinear structure defines two distinct magnetic sites with an average moment of 2.35(5) (=4.7(1)/2) $\mu_B$/Fe$^{2+}$, the two-\textbf{k} structure implies an ordered moment of 3.3(1) (= 4.7(1)/$\sqrt{2}$) $\mu_B$/Fe$^{2+}$ on all sites. Of these, only the latter structure yields a moment comparable to the spin-only localized moment of 4 $\mu_B$/Fe$^{2+}$, and is consistent with the observation of a single site Fe$^{2+}$ hyperfine split M\"{o}ssbauer spectrum.\cite{AFeX22122}

The noncollinear spin structure can be described as longitudinally polarized antiferromagnetic order on perpendicular Fe-O-Fe chains that intersect at oxygen sites. As for the ``2322" compounds, the intra-layer exchange interactions between the nearest and next nearest neighboring Fe sites involve three distinct exchange paths (green arrows, Fig. \ref{Neutron1}c): $J_1$ links spins within chains via the 180$^{\circ}$ Fe-O-Fe super-exchange interaction and is expected to be dominant. $J_2$ links perpendicular chains through the 90$^{\circ}$ Fe-O-Fe and the approximately 90$^{\circ}$ Fe-S-Fe super-exchange interactions.  $J_3$ corresponds to the Fe-S-Fe super-exchange with a bond angle of 100.22$^{\circ}$, which links parallel spin chains. The observed magnetic structure is favored by AFM $J_1$ and FM $J_3$ interactions. Isotropic $J_2$ interactions are frustrated due to the 90$^{\circ}$ angle between nearest neighbor spins.

\begin{figure}
  \includegraphics[width=\columnwidth]{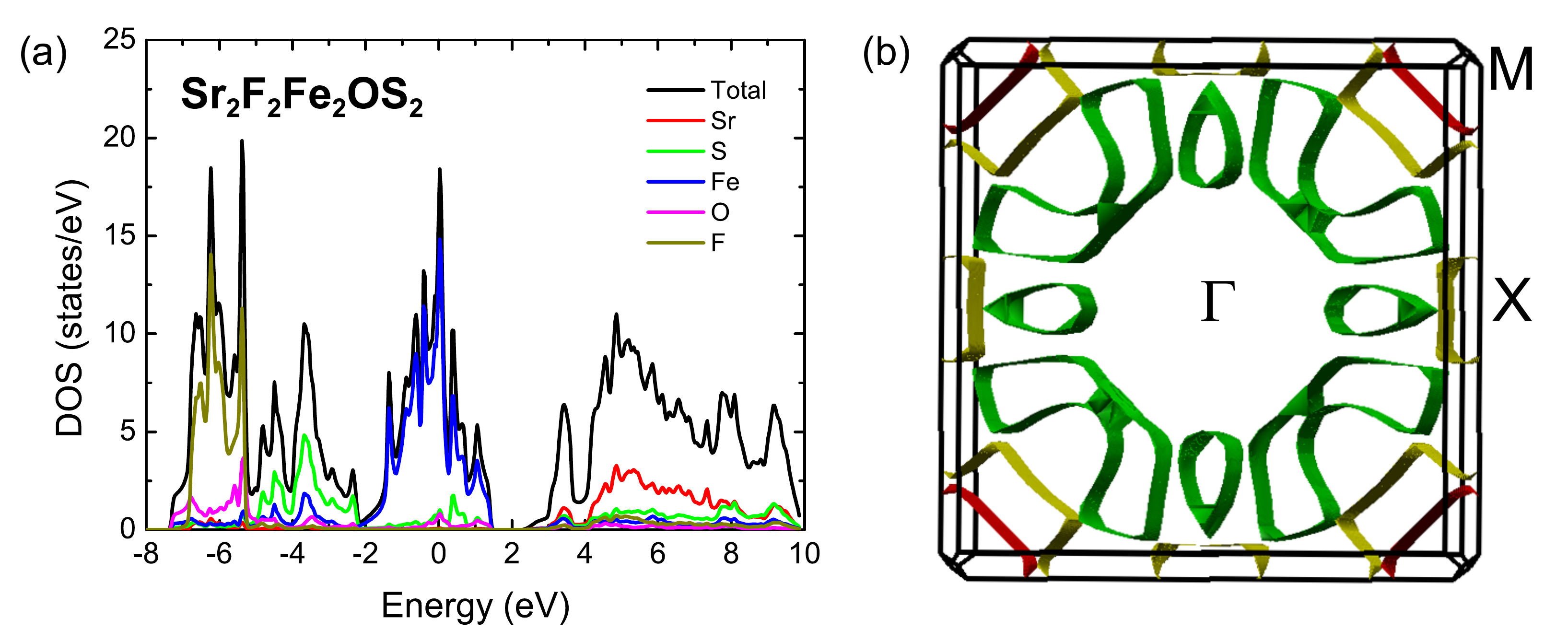}\\
  \caption{(a) Non-magnetic DOS and (b) Fermi surface of Sr$_2$F$_2$Fe$_2$OS$_2$, showing a narrow Fe 3$d$ band and quasi-two-dimensional features.}\label{DOSFS}
\end{figure}

The proposed magnetic structure differs from those inferred for the structurally analogous ``2322" systems La$_2$O$_3$T$_2$Se$_2$ (T = Mn, Fe, Co), for which the T = Mn compound shows a G-type AFM structure with all spins parallel to the $c$ axis,\cite{LaMnSe2322_ni} the T = Fe compound exhibits a bi-collinear AFM structure with all spins parallel to the $a$ axis,\cite{LaFeSe2322_neutron} while for T = Co a non-collinear plaquette AFM structure with 90$^{\circ}$ angles between the spin orientations of neighboring Co is observed.\cite{LaCoSe2322_neutron} None of these structures is consistent with the magnetic diffraction pattern for Sr$_2$F$_2$Fe$_2$OS$_2$ (Fig. 2c). The unusual diversity of magnetic structures indicates frustration in the magnetism of the checkerboard Fe$_2$OX$_2$ spin lattice.

\begin{table*}
  \setlength{\tabcolsep}{8pt}
  \caption{Relative energies $\Delta E$ (meV per unit cell) of six spin configurations and exchange constants $J_i$ (meV/bond) for different values of $U$ .}
  \begin{tabular}{ c || c| c | c | c | c | c || c | c | c }
  \hline
  $U$ (eV) & FM & AFM1 & AFM2 & AFM3 & AFM4 & AFM5 & $J_1$ & $J_2$ & $J_3$ \\ \hline
  1.5 & 0 & -191 & 121 & -64 & -319 & -247 & 27.4 & 11.1 & -12.3\\
  3.0 & 0 & -277 & -27 & -79 & -330 & -278 & 23.3 & 17.1 & -7.2\\
  4.5 & 0 & -248 & -36 & -73 & -262 & -228 & 18.0 & 15.2 & -5.5\\ \hline
    \end{tabular}
    \label{EJ}
\end{table*}

The lamellar magnetism revealed by the neutron data is also reflected in the thermodynamic properties. The DC magnetization $M(T)$, measured in a Quantum Design Magnetic Property Measurement System (QD MPMS) for $H$ = 1 T, shows a peak at $T_N$ = 106(2) K (Fig. \ref{MCR}a), as expected for long range AFM ordering. While the N\'{e}el temperature is consistent with a previous report,\cite{AFeX22122} the anomaly reported here is much sharper, indicating, as is generally the case in frustrated magnets, that sample quality can affect the transition. Heat capacity data (Fig. \ref{MCR}d), measured on a QD Physical Property Measurement System (PPMS)  also shows a sharp anomaly. By subtracting the estimated phonon contribution from a polynomial fit of the $C_{p}(T)$ data at temperatures away from $T_N$ (dashed line, Fig. \ref{MCR}d), the singular magnetic specific heat $C_m$ was extracted and plotted vs. $t=|T-T_{N}|/T_{N}$ in the inset of Fig. \ref{MCR}d. The observed logarithmic divergence of $C_m$ is expected for a two dimensional Ising system. Fig. \ref{MCR}b shows the temperature dependence of the sublattice magnetization $m(T)$ inferred from diffraction. Consistent with the specific heat, the data resemble the Onsager result for the square lattice Ising model (solid line).\cite{Onsager} Ising criticality is consistent with strongly anisotropic exchange interactions which also characterize the iron pnictides.\cite{Yildirim}

The zero-field-cooled (ZFC) and field-cooled (FC) magnetization curves (full and open symbols, respectively, in Fig. \ref{MCR}a) exhibit irreversibility below $T_N$ and this is also reflected by the magnetic hysteresis loop for $T$ = 2 K (inset, Fig. \ref{MCR}a). This indicates glassy behavior, consistent with the short range inter-plane correlations. While commensurate AFM order on a square lattice comes in two time-reversed versions, the spins in each of the perpendicular sets of chains can be separately reversed in the two-\textbf{k} structure, leading to a total of four different versions. Noting substantial in-plane correlations above $T_N$ indicated by the susceptibility maximum around 150 K (Fig.~\ref{MCR}), consider now the stacking  of such spin planes along \textbf{c}. To establish long range order along  \textbf{c}, Dzyaloshinskii-Moriya (DM) interactions between spins displaced by \textbf{c}/2 and exchange interactions between spins displaced by \textbf{c}, which - depending on the sign of the DM interaction - could even be frustrated, must select just one out of these four copies. Considering the slow dynamics associated with transitioning between the four different Ising spin planes, it is plausible the process fails due to kinetics and/or disorder.

Like all other ``2322" and ``22212" compounds, Sr$_2$F$_2$Fe$_2$OS$_2$ is an insulator for which the temperature dependent resistivity $\rho(T)$ (inset, Fig. \ref{MCR}c) can be described by $\rho\propto e^{E_{g}/k_{B}T}$, as shown in the Arrhenius plot (symbols, Fig. \ref{MCR}c). A linear fit of $\ln\rho$ vs. $1/T$ (solid line, Fig. \ref{MCR}c) suggests an activation gap $E_{g}$ = 0.28 eV, which is considerably larger than the previously reported value of 0.10 eV but equal to $E_{g}$ = 0.28 eV for Ba$_2$F$_2$Fe$_2$OSe$_2$.\cite{AFeX22122} At temperatures lower than 250 K, the resistivity starts to deviate from the activation gap model, which suggests the importance of hopping between localized gap states, as observed in LaMnPO$_{1-x}$F$_x$.\cite{LaMnPO}

To connect our experimental observations with the electronic structure of Sr$_2$F$_2$Fe$_2$OS$_2$, DFT calculations were performed using the full-potential linearized augmented plane wave (FP-LAPW) method implemented in the \textsc{wien2k} package.\cite{wien2k} The generalized gradient approximation (GGA) was used for the exchange correlation potential, and different effective onsite Coulomb repulsion values $U$ from 1.5 to 4.5 eV were considered. An Fe 3$d$ band between --2 eV and 1.5 eV is revealed in the non-magnetic density of states (DOS) spectrum (Fig. \ref{DOSFS}a). The 3$d$ band is narrower than that of LaOFeAs\cite{1111DFT} and BaFe$_2$As$_2$,\cite{122DFT} a direct consequence of the larger Fe square lattice in Sr$_2$F$_2$Fe$_2$OS$_2$ and the quasi-one-dimensional connectivity. Since the band width is proportional to the kinetic energy $t$, the narrow 3$d$ band suggests enhanced electron correlations $U/t$,  leading to the insulating behavior and a relatively high N\'{e}el temperature. The Sr-F layers (red and dark yellow lines, Fig. \ref{DOSFS}a) contribute negligible DOS to the Fermi level, hence the inter-layer electron hopping is strongly reduced, giving rise to the quasi-two-dimensional electronic structure. The low dimensionality of Sr$_2$F$_2$Fe$_2$OS$_2$ is also reflected in the Fermi surface, which mainly consists of quasi-2D ribbons extending along the $\Gamma-Z$ direction (Fig. \ref{DOSFS}b). 

The magnetic ground state and exchange constants $J_1$, $J_2$ and $J_3$ are estimated from spin-polarized calculations for the following six spin configurations (as shown in Ref. \onlinecite{LaFeX2322_PRL}): (1) FM, (2) checkerboard AFM (AFM1), (3) single stripe+FM (AFM2), (4) double stripe+FM (AFM3), (5) (1/2, 1/2) state with antiparallel alignment across the O sites (AFM4), (6) stripe AFM (AFM5). For different values of the Coulomb repulsion $U$, the ground state energies are listed in Table \ref{EJ}.

For $U$ = 0, finite DOS exists at the Fermi level for all spin configurations, while a Mott gap opens for finite $U$. Out of all the calculated spin configurations (Table \ref{EJ}), AFM4 is found to be the lowest energy state, with the ordered moment determined to be 3.49 $\mu_B$/Fe$^{2+}$. Although the specific orthogonal spin configuration is not uniquely identifiable in the DFT results due to software limitations, the moment value is in very good agreement with the value associated with the two-\textbf{k} magnetic structure (Fig. \ref{Neutron1}c). In addition, the signs of the exchange coupling constants $J_1$ and $J_2$ (Table \ref{EJ}) suggest an AFM-type Fe-O-Fe superexchange coupling and an FM-type coupling along the Fe-S-Fe paths, which are consistent with the two-\textbf{k} structure rather than the collinear structure with \textbf{k}$_{1}$ = (1/2, 0, 1/2). As suggested by the phase diagram proposed in Ref. \onlinecite{LaFeX2322_PRL}, the condition for AFM4 to be the ground state is $J_1 > J_2$ and $J_3 < 0$, which also agrees well with the calculated exchange constants of Sr$_2$F$_2$Fe$_2$OS$_2$. By calculating the energy difference between parallel and antiparallel spin configurations, the inter-layer interaction $J_{\perp}$ is estimated to be three orders of magnitude weaker.

In conclusion, we have provided evidence for a non-collinear magnetic structure in the layered oxychalcogenide Sr$_2$F$_2$Fe$_2$OS$_2$. The Fe spin lattice consists of orthogonally intersecting 1D AFM chains, with frustrated near neighbor interactions - a structure that to our knowledge has not previously been observed in a square lattice Mott insulator. The phase transition is of the 2D Ising variety and electronic two-dimensionality, interlayer magnetic frustration, and four degenerate versions of the two-\textbf{k} magnetic structure are contributing factors to magnetic hysteresis and short-range interlayer correlations.
The intersecting chain-like structure is the basis for the unusual non-collinear magnetism in Sr$_2$F$_2$Fe$_2$OS$_2$ and through effective electronic one-dimensionality perhaps also its Mott insulating state. The variety of magnetic structures suggests significant potential for tuning electronic correlations in the extended ``2322" family of materials.

\textbf{Acknowledgements}
The authors thank Quan Yin, Gabriel Kotliar, Rong Yu, Qimiao Si and Jian-Xin Zhu for useful discussions. The work at Rice University was supported by AFOSR-MURI. The work at IQM was supported by the U.S. Department of Energy (DOE), Office of Basic Energy Sciences, Division of Materials Sciences and Engineering under award DE-FG02-08ER46544.


\begin{thebibliography}{21}
\expandafter\ifx\csname natexlab\endcsname\relax\def\natexlab#1{#1}\fi
\expandafter\ifx\csname bibnamefont\endcsname\relax
  \def\bibnamefont#1{#1}\fi
\expandafter\ifx\csname bibfnamefont\endcsname\relax
  \def\bibfnamefont#1{#1}\fi
\expandafter\ifx\csname citenamefont\endcsname\relax
  \def\citenamefont#1{#1}\fi
\expandafter\ifx\csname url\endcsname\relax
  \def\url#1{\texttt{#1}}\fi
\expandafter\ifx\csname urlprefix\endcsname\relax\def\urlprefix{URL }\fi
\providecommand{\bibinfo}[2]{#2}
\providecommand{\eprint}[2][]{\url{#2}}

\bibitem[{\citenamefont{Mayer et~al.}(1992)\citenamefont{Mayer, Schneemeyer,
  Siegrist, Waszczak, and Dover}}]{LaFeX2322_mayer}
\bibinfo{author}{\bibfnamefont{J.~M.} \bibnamefont{Mayer}},
  \bibinfo{author}{\bibfnamefont{L.~E.} \bibnamefont{Schneemeyer}},
  \bibinfo{author}{\bibfnamefont{T.}~\bibnamefont{Siegrist}},
  \bibinfo{author}{\bibfnamefont{J.~V.} \bibnamefont{Waszczak}},
  \bibnamefont{and} \bibinfo{author}{\bibfnamefont{B.~V.} \bibnamefont{Dover}},
  \bibinfo{journal}{Angelic. Chem. Int. Ed. Engl.}
  \textbf{\bibinfo{volume}{31}}, \bibinfo{pages}{1645} (\bibinfo{year}{1992}).

\bibitem[{\citenamefont{Zhu et~al.}(2010)\citenamefont{Zhu, Yu, Wang, Zhao,
  Jones, Dai, Abrahams, Morosan, Fang, and Si}}]{LaFeX2322_PRL}
\bibinfo{author}{\bibfnamefont{J.-X.} \bibnamefont{Zhu}},
  \bibinfo{author}{\bibfnamefont{R.}~\bibnamefont{Yu}},
  \bibinfo{author}{\bibfnamefont{H.}~\bibnamefont{Wang}},
  \bibinfo{author}{\bibfnamefont{L.~L.} \bibnamefont{Zhao}},
  \bibinfo{author}{\bibfnamefont{M.~D.} \bibnamefont{Jones}},
  \bibinfo{author}{\bibfnamefont{J.}~\bibnamefont{Dai}},
  \bibinfo{author}{\bibfnamefont{E.}~\bibnamefont{Abrahams}},
  \bibinfo{author}{\bibfnamefont{E.}~\bibnamefont{Morosan}},
  \bibinfo{author}{\bibfnamefont{M.}~\bibnamefont{Fang}}, \bibnamefont{and}
  \bibinfo{author}{\bibfnamefont{Q.}~\bibnamefont{Si}}, \bibinfo{journal}{Phys.
  Rev. Lett.} \textbf{\bibinfo{volume}{104}}, \bibinfo{pages}{216405}
  (\bibinfo{year}{2010}).

\bibitem[{\citenamefont{Wang et~al.}(2010)\citenamefont{Wang, Tan, Feng, Ma,
  Jiang, Xu, Cao, Matsubayashi, and Uwatoko}}]{LaCoSe2322_JACS}
\bibinfo{author}{\bibfnamefont{C.}~\bibnamefont{Wang}},
  \bibinfo{author}{\bibfnamefont{M.~Q.} \bibnamefont{Tan}},
  \bibinfo{author}{\bibfnamefont{C.~M.} \bibnamefont{Feng}},
  \bibinfo{author}{\bibfnamefont{Z.~F.} \bibnamefont{Ma}},
  \bibinfo{author}{\bibfnamefont{S.}~\bibnamefont{Jiang}},
  \bibinfo{author}{\bibfnamefont{Z.~A.} \bibnamefont{Xu}},
  \bibinfo{author}{\bibfnamefont{G.~H.} \bibnamefont{Cao}},
  \bibinfo{author}{\bibfnamefont{K.}~\bibnamefont{Matsubayashi}},
  \bibnamefont{and} \bibinfo{author}{\bibfnamefont{Y.}~\bibnamefont{Uwatoko}},
  \bibinfo{journal}{J. Am. Chem. Soc.} \textbf{\bibinfo{volume}{132}},
  \bibinfo{pages}{7069} (\bibinfo{year}{2010}).

\bibitem[{\citenamefont{Fuwa et~al.}(2010{\natexlab{a}})\citenamefont{Fuwa,
  Wakeshima, and Hinatsu}}]{LaCoSe2322_Fuwa}
\bibinfo{author}{\bibfnamefont{Y.}~\bibnamefont{Fuwa}},
  \bibinfo{author}{\bibfnamefont{M.}~\bibnamefont{Wakeshima}},
  \bibnamefont{and} \bibinfo{author}{\bibfnamefont{Y.}~\bibnamefont{Hinatsu}},
  \bibinfo{journal}{Solid State Comm.} \textbf{\bibinfo{volume}{150}},
  \bibinfo{pages}{1698} (\bibinfo{year}{2010}{\natexlab{a}}).

\bibitem[{\citenamefont{Ni et~al.}(2010)\citenamefont{Ni, Climent-Pascual, Jia,
  Huang, and Cava}}]{LaMnSe2322_ni}
\bibinfo{author}{\bibfnamefont{N.}~\bibnamefont{Ni}},
  \bibinfo{author}{\bibfnamefont{E.}~\bibnamefont{Climent-Pascual}},
  \bibinfo{author}{\bibfnamefont{S.}~\bibnamefont{Jia}},
  \bibinfo{author}{\bibfnamefont{Q.}~\bibnamefont{Huang}}, \bibnamefont{and}
  \bibinfo{author}{\bibfnamefont{R.~J.} \bibnamefont{Cava}},
  \bibinfo{journal}{Phys. Rev. B} \textbf{\bibinfo{volume}{82}},
  \bibinfo{pages}{214419} (\bibinfo{year}{2010}).

\bibitem[{\citenamefont{Liu et~al.}(2011)\citenamefont{Liu, Zhang, Cheng, Luo,
  Ying, Yan, Zhang, Wang, Xiang, Ye et~al.}}]{LaMnSe2322_BaMnSe22122_XHChen}
\bibinfo{author}{\bibfnamefont{R.~H.} \bibnamefont{Liu}},
  \bibinfo{author}{\bibfnamefont{J.~S.} \bibnamefont{Zhang}},
  \bibinfo{author}{\bibfnamefont{P.}~\bibnamefont{Cheng}},
  \bibinfo{author}{\bibfnamefont{X.~G.} \bibnamefont{Luo}},
  \bibinfo{author}{\bibfnamefont{J.~J.} \bibnamefont{Ying}},
  \bibinfo{author}{\bibfnamefont{Y.~J.} \bibnamefont{Yan}},
  \bibinfo{author}{\bibfnamefont{M.}~\bibnamefont{Zhang}},
  \bibinfo{author}{\bibfnamefont{A.~F.} \bibnamefont{Wang}},
  \bibinfo{author}{\bibfnamefont{Z.~J.} \bibnamefont{Xiang}},
  \bibinfo{author}{\bibfnamefont{G.~J.} \bibnamefont{Ye}},
  \bibnamefont{et~al.}, \bibinfo{journal}{Phys. Rev. B}
  \textbf{\bibinfo{volume}{83}}, \bibinfo{pages}{174450}
  (\bibinfo{year}{2011}).

\bibitem[{\citenamefont{Free et~al.}(2011)\citenamefont{Free, Withers, Hickey,
  and Evans}}]{RTSe2322}
\bibinfo{author}{\bibfnamefont{D.~G.} \bibnamefont{Free}},
  \bibinfo{author}{\bibfnamefont{N.~D.} \bibnamefont{Withers}},
  \bibinfo{author}{\bibfnamefont{P.~J.} \bibnamefont{Hickey}},
  \bibnamefont{and} \bibinfo{author}{\bibfnamefont{J.~S.~O.}
  \bibnamefont{Evans}}, \bibinfo{journal}{Chem. Mater.}
  \textbf{\bibinfo{volume}{23}}, \bibinfo{pages}{1625} (\bibinfo{year}{2011}).

\bibitem[{\citenamefont{Charkina et~al.}(2011)\citenamefont{Charkina,
  Plotnikov, Sadakov, Omel'yanovskii, and Kazakov}}]{RFeS2322}
\bibinfo{author}{\bibfnamefont{D.~O.} \bibnamefont{Charkina}},
  \bibinfo{author}{\bibfnamefont{V.~A.} \bibnamefont{Plotnikov}},
  \bibinfo{author}{\bibfnamefont{A.~V.} \bibnamefont{Sadakov}},
  \bibinfo{author}{\bibfnamefont{O.~E.} \bibnamefont{Omel'yanovskii}},
  \bibnamefont{and} \bibinfo{author}{\bibfnamefont{S.~M.}
  \bibnamefont{Kazakov}}, \bibinfo{journal}{J. Alloys. and Comp.}
  \textbf{\bibinfo{volume}{509}}, \bibinfo{pages}{7344} (\bibinfo{year}{2011}).

\bibitem[{\citenamefont{Ni et~al.}(2011)\citenamefont{Ni, Jia, Huang,
  Climent-Pascual, and Cava}}]{RFeSe2322}
\bibinfo{author}{\bibfnamefont{N.}~\bibnamefont{Ni}},
  \bibinfo{author}{\bibfnamefont{S.}~\bibnamefont{Jia}},
  \bibinfo{author}{\bibfnamefont{Q.}~\bibnamefont{Huang}},
  \bibinfo{author}{\bibfnamefont{E.}~\bibnamefont{Climent-Pascual}},
  \bibnamefont{and} \bibinfo{author}{\bibfnamefont{R.~J.} \bibnamefont{Cava}},
  \bibinfo{journal}{Phys. Rev. B} \textbf{\bibinfo{volume}{83}},
  \bibinfo{pages}{224403} (\bibinfo{year}{2011}).

\bibitem[{\citenamefont{Free and Evans}(2010)}]{LaFeSe2322_neutron}
\bibinfo{author}{\bibfnamefont{D.~G.} \bibnamefont{Free}} \bibnamefont{and}
  \bibinfo{author}{\bibfnamefont{J.~S.~O.} \bibnamefont{Evans}},
  \bibinfo{journal}{Phys. Rev. B} \textbf{\bibinfo{volume}{81}},
  \bibinfo{pages}{214433} (\bibinfo{year}{2010}).

\bibitem[{\citenamefont{Fuwa et~al.}(2010{\natexlab{b}})\citenamefont{Fuwa,
  Endo, Wakeshima, Hinatsu, and Ohoyama}}]{LaCoSe2322_neutron}
\bibinfo{author}{\bibfnamefont{Y.}~\bibnamefont{Fuwa}},
  \bibinfo{author}{\bibfnamefont{T.}~\bibnamefont{Endo}},
  \bibinfo{author}{\bibfnamefont{M.}~\bibnamefont{Wakeshima}},
  \bibinfo{author}{\bibfnamefont{Y.}~\bibnamefont{Hinatsu}}, \bibnamefont{and}
  \bibinfo{author}{\bibfnamefont{K.}~\bibnamefont{Ohoyama}},
  \bibinfo{journal}{J. Am. Chem. Soc.} \textbf{\bibinfo{volume}{132}},
  \bibinfo{pages}{18020} (\bibinfo{year}{2010}{\natexlab{b}}).

\bibitem[{\citenamefont{Kabbour et~al.}(2008)\citenamefont{Kabbour, Janod,
  Corraze, Danot, Lee, Whangbo, and Cario}}]{AFeX22122}
\bibinfo{author}{\bibfnamefont{H.}~\bibnamefont{Kabbour}},
  \bibinfo{author}{\bibfnamefont{E.}~\bibnamefont{Janod}},
  \bibinfo{author}{\bibfnamefont{B.}~\bibnamefont{Corraze}},
  \bibinfo{author}{\bibfnamefont{M.}~\bibnamefont{Danot}},
  \bibinfo{author}{\bibfnamefont{C.}~\bibnamefont{Lee}},
  \bibinfo{author}{\bibfnamefont{M.-H.} \bibnamefont{Whangbo}},
  \bibnamefont{and} \bibinfo{author}{\bibfnamefont{L.}~\bibnamefont{Cario}},
  \bibinfo{journal}{J. Am. Chem. Soc.} \textbf{\bibinfo{volume}{130}},
  \bibinfo{pages}{8261} (\bibinfo{year}{2008}).

\bibitem[{\citenamefont{Onsager}(1944)\citenamefont{Onsager}}]{Onsager}
\bibinfo{author}{\bibfnamefont{L.}~\bibnamefont{Onsager}},
  \bibinfo{journal}{Phys. Rev.} \textbf{\bibinfo{volume}{65}},
  \bibinfo{pages}{117} (\bibinfo{year}{1944}).


\bibitem[{\citenamefont{Yildirim}(2008)\citenamefont{Yildirim}}]{Yildirim}
\bibinfo{author}{\bibfnamefont{T.}~\bibnamefont{Yildirim}},
  \bibinfo{journal}{Phys. Rev. Lett.} \textbf{\bibinfo{volume}{101}},
  \bibinfo{pages}{057010} (\bibinfo{year}{2008}).

\bibitem[{\citenamefont{He et~al.}(2011)\citenamefont{He, Wang, Shi, Yang, Li,
  and Chen}}]{Na2Fe2OSe2}
\bibinfo{author}{\bibfnamefont{J.~B.} \bibnamefont{He}},
  \bibinfo{author}{\bibfnamefont{D.~M.} \bibnamefont{Wang}},
  \bibinfo{author}{\bibfnamefont{H.~L.} \bibnamefont{Shi}},
  \bibinfo{author}{\bibfnamefont{H.~X.} \bibnamefont{Yang}},
  \bibinfo{author}{\bibfnamefont{J.~Q.} \bibnamefont{Li}}, \bibnamefont{and}
  \bibinfo{author}{\bibfnamefont{G.~F.} \bibnamefont{Chen}},
  \bibinfo{journal}{Phys. Rev. B} \textbf{\bibinfo{volume}{84}},
  \bibinfo{pages}{205212} (\bibinfo{year}{2011}).

\bibitem[{\citenamefont{Rodriguez-Carvajal}(1993)}]{FullProf}
\bibinfo{author}{\bibfnamefont{J.}~\bibnamefont{Rodriguez-Carvajal}},
  \bibinfo{journal}{Physica B} \textbf{\bibinfo{volume}{192}},
  \bibinfo{pages}{55} (\bibinfo{year}{1993}).

\bibitem[{\citenamefont{SuppMat}()}]{SuppMat}
\bibinfo{title}{See supplemental material at [url will be inserted by
  publisher] for the refined atomic positions and isotropic thermal factors of Sr$_2$F$_2$Fe$_2$OS$_2$.}

\bibitem[{\citenamefont{Wills}(2000)}]{SARAh}
\bibinfo{author}{\bibfnamefont{A.~S.} \bibnamefont{Wills}},
  \bibinfo{journal}{Physica B} \textbf{\bibinfo{volume}{276-278}},
  \bibinfo{pages}{680} (\bibinfo{year}{2000}).

\bibitem[{\citenamefont{Kovalev}(1965)}]{Kovalev}
\bibinfo{author}{\bibfnamefont{O.~V.} \bibnamefont{Kovalev}},
  \emph{\bibinfo{title}{Irreducible Representations of the Space Groups}}
  (\bibinfo{publisher}{Gordon and Breach, New York}, \bibinfo{year}{1965}).

\bibitem[{\citenamefont{Warren}(1941)}]{Warren}
\bibinfo{author}{\bibfnamefont{B.~E.} \bibnamefont{Warren}},
  \bibinfo{journal}{Phys. Rev.} \textbf{\bibinfo{volume}{59}},
  \bibinfo{pages}{693} (\bibinfo{year}{1941}).

\bibitem[{\citenamefont{Simonson et~al.}(2011)\citenamefont{Simonson, Post,
  Marques, Smith, Khatib, Basov, and Aronson}}]{LaMnPO}
\bibinfo{author}{\bibfnamefont{J.~W.} \bibnamefont{Simonson}},
  \bibinfo{author}{\bibfnamefont{K.}~\bibnamefont{Post}},
  \bibinfo{author}{\bibfnamefont{C.}~\bibnamefont{Marques}},
  \bibinfo{author}{\bibfnamefont{G.}~\bibnamefont{Smith}},
  \bibinfo{author}{\bibfnamefont{O.}~\bibnamefont{Khatib}},
  \bibinfo{author}{\bibfnamefont{D.~N.} \bibnamefont{Basov}}, \bibnamefont{and}
  \bibinfo{author}{\bibfnamefont{M.~C.} \bibnamefont{Aronson}},
  \bibinfo{journal}{Phys. Rev. B} \textbf{\bibinfo{volume}{84}},
  \bibinfo{pages}{165129} (\bibinfo{year}{2011}).

\bibitem[{\citenamefont{Blaha et~al.}()\citenamefont{Blaha, Schwarz, Madsen,
  Kvasnicka, and Luitz}}]{wien2k}
\bibinfo{author}{\bibfnamefont{P.~K.} \bibnamefont{Blaha}},
  \bibinfo{author}{\bibfnamefont{K.}~\bibnamefont{Schwarz}},
  \bibinfo{author}{\bibfnamefont{G.}~\bibnamefont{Madsen}},
  \bibinfo{author}{\bibfnamefont{D.}~\bibnamefont{Kvasnicka}},
  \bibnamefont{and} \bibinfo{author}{\bibfnamefont{J.}~\bibnamefont{Luitz}},
  \bibinfo{journal}{\textsc{WIEN2K} package, http://www.wien2k.at}.

\bibitem[{\citenamefont{Singh and Du}(2008)}]{1111DFT}
\bibinfo{author}{\bibfnamefont{D.~J.} \bibnamefont{Singh}} \bibnamefont{and}
  \bibinfo{author}{\bibfnamefont{M.-H.} \bibnamefont{Du}},
  \bibinfo{journal}{Phys. Rev. Lett.} \textbf{\bibinfo{volume}{100}},
  \bibinfo{pages}{237003} (\bibinfo{year}{2008}).

\bibitem[{\citenamefont{Shein and Ivanovskii}(2008)}]{122DFT}
\bibinfo{author}{\bibfnamefont{I.~R.} \bibnamefont{Shein}} \bibnamefont{and}
  \bibinfo{author}{\bibfnamefont{A.~L.} \bibnamefont{Ivanovskii}},
  \bibinfo{journal}{JETP Lett.} \textbf{\bibinfo{volume}{88}},
  \bibinfo{pages}{107} (\bibinfo{year}{2008}).

\end{thebibliography}

\end{document}